\documentclass [11pt, letterpaper]{article}
\usepackage{fullpage}
\usepackage{amsmath}
\usepackage{amssymb}
\usepackage{graphicx}
\usepackage{mathrsfs}
\usepackage{wrapfig}
\usepackage{boxedminipage}
\usepackage{epsfig}
\usepackage{setspace}
\usepackage{subfigure}
\usepackage{float}
\usepackage{hyperref}
\usepackage{soul}
\usepackage{enumerate}

\usepackage{cite}
\newcommand{\be}{\begin{equation}}
\newcommand{\bea}{\begin{eqnarray}}
\newcommand{\eea}{\end{eqnarray}}
\newcommand{\ba}{\begin{align}}
\newcommand{\ea}{\end{align}}
\newcommand{\ee}{\end{equation}}
\newcommand{\nn}{\nonumber}

\newcommand{\CO}{{\cal{O}}}
\begin{document}

\begin{titlepage}
\vspace{10mm}
\begin{flushright}
 IPM/P-2013/035 \\
\end{flushright}
  %
\vspace*{20mm}
\begin{center}
{\Large {\bf  Entanglement Entropy for Logarithmic Conformal Field Theory }\\
}

\vspace*{15mm}
\vspace*{1mm}
{Mohsen Alishahiha${}^a$, Amin Faraji Astaneh$^{b,c}$ 
and  M. Reza Mohammadi Mozaffar$^a$ }

 \vspace*{1cm}

{\it ${}^a$ School of Physics, Institute for Research in Fundamental Sciences (IPM)\\
P.O. Box 19395-5531, Tehran, Iran \\  
${}^b$ Department of Physics, Sharif University of Technology,\\
P.O. Box 11365-9161, Tehran, Iran\\
${}^c $ School of Particles and Accelerators\\Institute for Research in Fundamental Sciences (IPM)\\
P.O. Box 19395-5531, Tehran, Iran \\ 
}
 \vspace*{0.5cm}
{E-mails: {\tt alishah@ipm.ir, faraji@ipm.ir, m$_{-}$mohammadi@ipm.ir}}%

\vspace*{2cm}
\end{center}

\begin{abstract}
We study holographic entanglement entropy for certain logarithmic conformal field 
theories by making use of their gravity descriptions. The corresponding gravity
descriptions are  provided by higher derivative gravity at critical points where the
equations of motion degenerate leading to a log gravity.
 When a central charge of the dual theory is zero, the entanglement entropy has 
a new divergent term whose coefficient is given by the ``new anomaly'' of the 
logarithmic conformal field theory.
\end{abstract}

\end{titlepage}

\section{Introduction}
In this paper using the holographic description of  entanglement entropy
\cite{{RT:2006PRL},{RT:2006}} we study  entanglement entropy for certain  logarithmic conformal
field theories (LCFT's)\cite{Gurarie:1993xq}. To do so, we utilize  gravity descriptions of  LCFT's
which may be provided by higher derivative  gravities at  critical points. In three dimensions
such theories are given by  
Topologically Massive Gravity (TMG) \cite{{Deser:1981wh},{Deser:1982vy}} or 
New Massive Gravity (NMG) \cite{Bergshoeff:2009hq} while for higher dimensions the theories are  known
as log gravity\cite{{Deser:2011xc},{Alishahiha:2011},{Gullu:2011sj},{Bergshoeff:2011ri}}. 

Since in a gravity with  higher derivative terms the corresponding equations of motion are higher
order differential equations, there is a possibility to have  critical points where  the equations 
of motion degenerate leading  to a logarithmic solution\footnote{
The fact that at the critical point there is a new logarithmic mode,
has  been first obtained for  TMG in  \cite{Grumiller:2008qz} where the authors have also conjectured 
that the dual field theory must be a LCFT.}. In general the resultant logarithmic solutions have 
the following form \cite{Alishahiha:2011}
\be\label{LOGS}
ds^2_{d+1}=\frac{L^2}{r^2}\left[-F(x_+,r) dx_+^2-2 dx_+dx_-+d\vec{x}^2+dr^2\right],
\ee
with
\be
F(x_+,r)=\beta_1(x_+)+\beta_2(x_+)r^{d-1}+\left[\beta_3(x_+)+\beta_4(x_+)r^{d-1}\right] \ln\left(\frac{r}{L}\right),
\ee
where $\beta_i$'s are arbitrary functions of $x_+$.

More precisely for $\beta_3\neq 0$  at the critical points these higher derivative gravities  admit a new vacuum solution\cite{{Alishahiha:2011},{AyonBeato:2004fq},{AyonBeato:2009yq}} which is not asymptotically locally AdS. To accommodate this solution one needs to change the asymptotic behaviour of the AdS metric. Actually using the Fefferman-Graham 
coordinates for a $d+1$ dimensional  metric
\be
ds^2_{d+1}=\frac{d\rho^2}{4\rho^2}+\frac{1}{\rho}\;g_{ij}(\rho,\vec{x})\;dx^idx^j,
\ee
the equations of motion allow us to have a wider class of the boundary conditions for the metric as follows
\cite{{Grumiller:2008es},{Skenderis:2009nt},{Alishahiha:2010bw}}
\be
g_{ij}(\rho,\vec{x})=b_{(0)ij}(\vec{x})\ln\rho+g_{(0)ij}(\vec{x})+\cdots+\bigg(b_{(d)ij}(\vec{x})\ln\rho+g_{(d)ij}(\vec{x})\bigg)\rho^d+\cdots.
\ee
It is then obvious that for $b_{(0)ij}\neq 0$ the solution is not asymptotically locally AdS. Indeed in this case in order to  maintain the variational principle well posed with the Dirichlet boundary condition one needs to modify the variational principle by imposing an additional boundary condition. In fact, from the above asymptotic expansion one has
\be
g_{(0)ij}=\lim_{\rho\rightarrow 0}(g_{ij}-\rho\;\partial_\rho g_{ij}),\;\;\;\;\;\;\;\;\;
b_{(0)ij}=\lim_{\rho\rightarrow 0}\rho\;\partial_\rho g_{ij},
\ee
which shows that the boundary condition can be fixed not only by the value of the boundary metric but also by its radial derivative. 

From AdS/CFT correspondence \cite{Maldacena:1997re} point of view this means that in the dual CFT there are two operators which may be associated with the metric, one of them is sourced by  $b_{(0)ij}$    and the other by $g_{(0)ij}$ . This is, indeed, the reason why the corresponding dual field theory
might be a LCFT \cite{{Grumiller:2008qz},{Skenderis:2009nt}} ( see also\cite{{Grumiller:2009mw},{Grumiller:2009sn},{Alishahiha:2010bw}}).
It is important to mention that in this context adding  log term to an AdS solution may be holographically 
identified to a deformation of the dual CFT with an irrelevant operator. Therefore, adding this term would destroy the conformal symmetry at UV
and it is not clear how to apply the AdS/CFT correspondence. Nevertheless, following \cite{Skenderis:2009nt} we will assume that the deformation  is sufficiently small and this term may be treated perturbatively. 
To conclude, it is believe that  the dual field theory of higher derivative gravity on the logarithmic
solution \eqref{LOGS} is a LCFT.

Having established  gravity descriptions for LCFT's it is then natural to study different 
aspects of them by making use of AdS/CFT correspondence. In particular one may study
 entanglement entropy in a LCFT using the holographic description 
of the entanglement entropy\cite{{RT:2006PRL},{RT:2006}}\footnote{The importance of computing
entanglement entropy for LCFT's has also been mentioned in\cite{Grumiller:2013at}. We would like to thank 
D. Grumiller and  T. Zojer for bringing this paper to our attentions.}.
We note, however, that since in the case of our interest the corresponding  action 
of the dual gravity  
contains higher derivative terms the simple expression of the holographic entanglement
entropy in terms of a minimal surface in the bulk is  not applicable.
Thus one needs to proceed with  another procedure.

Of course  unlike the Wald formula \cite{Wald:1993nt} for the black hole entropy, there is no
such a generalization for the entanglement entropy for an arbitrary action with higher 
derivative terms. The only case where the minimization procedure is argued to work 
is for that of Lovelock theories\cite{{Hung:2011xb},{deBoer:2011wk}}. We note, however, that
neither TMG and NMG  nor higher dimensional log gravities belong to this category.  

Therefore to compute the holographic entanglement entropy one should proceed with another
method.  Actually  using the method based on a regularization of squashed cones introduced 
in \cite{Fursaev:2013fta}, the authors of \cite{Bhattacharyya:2013gra} have 
computed the entanglement entropy of an AdS vacuum in NMG  as well as  a five dimensional higher 
derivative gravity away from
the critical point where they have found that the  regularization  procedure  leads to the expected 
universal terms in the entanglement entropy for spherical and cylindrical entangling surfaces
\footnote{An alternative way to compute the entanglement entropy for arbitrary gravitational model is
the procedure presented in\cite{Lewkowycz:2013nqa} where the authors introduced the generalized gravitational entropy.}.
 In the present  paper we will extend this consideration to the model at the critical points where the vacuum is not an AdS solution\footnote{Holographic entanglement entropy for a solution
which is not asymptotically AdS has been also considered in\cite{Anninos:2013nja}.}.

The paper is organized as follows. In the next section we will study holographic entanglement
entropy for NMG model at the critical point. In section three we shall redo 
the same calculations for higher dimensional higher derivative  log gravities. The last section is devoted to discussions.


\section{Three dimensional log-gravity}

In this section we shall consider two dimensional LCFT's whose gravity duals may be provided by NMG or
TMG at the critical points. To explore the physical content of these models, it is worth 
mentioning that  TMG model is parity odd while the NMG model is parity even. From dual 
field theory point of view this property leads to the fact that the LCFT dual to  TMG has logarithmic behaviour in the
left hand sector while the right hand sector is usual CFT, while for LCFT dual to NMG both sectors are 
logarithmic.

 More precisely in the case of TMG the dual LCFT has central charges $c_L=0, c_R=\frac{3L}{G_N}$
and new anomaly $b_R=0$, $b_L=\frac{-3L}{G_N}$, while for NMG one has $c_L=c_R=0$ and  $b_R=b_L=-\frac{12L}{G_N}$. Here $G_N$ is the Newton constant and $L$ is the radius of space-time.

In  what follows we shall study entanglement entropy of  a two dimensional LCFT whose gravitational description is provided by NMG at the critical point.  To do so, we will consider an interval with the 
width $\ell$ as the entangling region. 
We will back to the TMG model in the
discussions section.

The action  of NMG model is 
\begin{equation}
S_{NMG}=-\frac{1}{16\pi G_N}\int d^3x \sqrt{g}\left[R-2\Lambda-
\frac{1}{m^2}\left(R_{\mu\nu}R^{\mu\nu}-\frac{3}{8}R^2\right)\right]\ .
\end{equation}
For generic $m$ this model  admits several vacua, including an AdS$_3$ vacuum. It is believed that the NMG model on an asymptotically locally AdS$_3$ solution 
may have a dual CFT whose central charges are given by\cite{{Bergshoeff:2009aq},{Liu:2009kc}}
\be
c_L=c_R=\frac{3L}{2G_N}\left(1-\frac{1}{2m^2L^2}\right),
\ee
where $L$ is the radius of the AdS solution. At critical point where $m^2L^2=\frac{1}{2}$ both central charges vanish and indeed the model exhibits a logarithmic solution.

Entanglement entropy for a CFT dual to an asymptotically locally AdS solution in NMG model
has  been recently studied  in\cite{Bhattacharyya:2013gra}.  As we have already mentioned 
when the action contains higher derivative terms the holographic description of entanglement entropy 
in terms of  a  minimal surface does not work. Nevertheless to compute the entanglement entropy for the present case,
based on the results of  \cite{Fursaev:2013fta}, the authors of \cite{Bhattacharyya:2013gra} have 
considered the following entropy functional to be minimized  to compute the entanglement entropy
\be\label{EN}
S_{EE}=\frac{1}{4G_N}\int dx\sqrt{g_{\rm ind}}\left\{1-\frac{1}{m^2}\left[\left(R_{\mu\nu}n^\mu_in^\nu_i-
\frac{1}{2}K_i^2\right)-\frac{3}{4}R\right]\right\},
\ee
where $i=1,2$ denotes two transverse directions to a co-dimension two hypersurface in the bulk,
$n_i^\mu$  are two unit  mutually orthogonal normal vectors on the co-dimension two hypersurface 
and $K^{(i)}$ is trace of the two extrinsic curvature tensors defined by
\be 
K_{\mu\nu}^{(i)}=\pi^\sigma_{\ \mu} \pi^\rho_{\ \nu}\nabla_\rho(n_i)_\sigma,\;\;\;\;\;\;\;
{\rm with}\;\;\;\;  \pi^\sigma_{\ \mu}=\epsilon^\sigma_{\ \mu}+\xi\sum_{i=1,2}(n_i)^\sigma(n_i)_\mu\ ,
\ee
where $\xi=-1$ for space-like and $\xi=1$ for time-like vectors.

In our notation for the AdS geometry 
\be
ds^2=\frac{L^2}{r^2}(-dt^2+dx^2+dr^2),
\ee
the co-dimension two hypersurface in the bulk is fixed by $x=x(r)$ and $t=0$. This 
parametrization can be used to compute the entanglement entropy of an interval with the width 
$\ell$ along $x$ direction in the dual conformal field theory. Indeed plugging this parametrization into the 
entropy functional \eqref{EN} and minimizing it, one may find  the profile of the co-dimension two hypersurface\cite{Bhattacharyya:2013gra}, $x(r)=\sqrt{\ell^2/4-r^2}$. Then, this can be used
to compute the entropy functional on this hypersurface which is indeed the  entanglement entropy. 
Doing so, one finds the universal part of the
entanglement entropy as follows\cite{Bhattacharyya:2013gra}
 \be
S_{EE}=\frac{ L}{4G_N}\left(1-\frac{1}{2m^2L^2}\right)\ln\frac{\ell}{\epsilon},
\ee
where $\epsilon$ is a UV cut off. From this expression it is evident that at the critical point where $m^2L^2=\frac{1}{2}$ the universal
term vanishes. Of course it was expected due to fact that both central charges are zero  at 
the critical point.  It is then interesting to study entanglement entropy at this point.

As we already mentioned at the critical point the model admits a new solution.  It is then 
important to compute the holographic entanglement entropy for this background. The 
corresponding background is
\be\label{logm}
ds^2=\frac{L^2}{r^2}\left[-\beta \ln\left(\frac{r}{L}\right) dx_+^2-2 dx_+dx_-+dr^2\right], 
\ee
 where the light-cone coordinates are defined as $x_\pm=\frac{t\pm x}{\sqrt{2}}$.

 Note that in this metric the constant parameter $\beta$ in the logarithmic term can be set to one by a rescaling of the coordinates. Nevertheless in order to trace effects of the 
logarithmic term we keep it in the metric. More importantly, as we already mentioned, the above metric
is not asymptotically AdS. Actually, its deviation from an AdS geometry is holographically
identified to  deforming the dual CFT with an irrelevant operator. On the other hand  in order to 
work within the framework of a CFT where one can benefit from the power of conformal symmetry,
 one needs to assume that the coefficient of the irrelevant operator is sufficiently small. 
From gravity point of view this,  in turn, indicates that $\beta$ is very small. Therefore in what 
follows we will minimize the entropy functional \eqref{EN}  for the
logarithmic metric \eqref{logm} with the 
assumption that $\beta$ is very small. 
 
To proceed we should consider a co-dimension two hypersurface in the bulk whose boundary 
coincides with the boundary of the entangling region. The corresponding co-dimension two hypersurface is
parametrized as follows
\begin{equation}\label{two}
x_++x_-=0\quad , \quad x_+-x_-=\sqrt{2}f(r)\ .
\end{equation}
Then the induced metric reads
\begin{eqnarray}\label{induced}
ds^2=\frac{{L}^2}{r^2}\left[1+f'^2(r)-\frac{\beta}{2}f'^2(r)\ln\left( \frac{r}{{L}}\right)\right]dr^2.
\end{eqnarray}
The corresponding two unit vectors are given by
\begin{eqnarray}\label{sigma}
&&\hspace*{0.5cm}x_++x_-=\text{const.}\hspace*{2.24cm}n_1=\frac{L}{r\sqrt{2-\beta\ln\left(\frac{r}{L}\right)}}(0,1,1)\ ,\nonumber\\
&&\hspace*{0.5cm}x_+-x_--\sqrt{2}f(r)=\text{const.}\hspace*{0.5cm}n_2=\frac{L}{r\sqrt{
2+\beta\ln\left(\frac{r}{L}\right)+2f'^2
}}(-\sqrt{2}f',1,-1).
\end{eqnarray}
Using these expressions one can compute the extrinsic curvatures whose
traces are found to be 
\begin{eqnarray}\label{tracek}
K_{1}&=&-\frac{\beta \left(-4+(\beta-4) \ln\frac{r}{{L}}+2 \beta \ln^2 \frac{r}{{L}}\right) f'}{\sqrt{2} {L} \left(-2+\beta \ln \frac{r}{{L}}\right)^{3/2} \left(2+\beta \ln \frac{r}{{L}}+2 f'^2\right)}\ ,\nonumber\\ &&\cr
K_{2}&=&\frac{\sqrt{2} \left(\left(-2 (2+\beta)+\beta^2 \ln^2\frac{r}{{L}}\right) f'+\left(-4-\beta+2 \beta \ln \frac{r}{{L}}\right) f'^3+z \left(4-\beta^2 \ln^2\frac{r}{{L}}\right) f''\right)}{{L} \left(-2+\beta \ln \frac{r}{{L}}\right) \left(2+\beta \ln \frac{r}{{L}}+2 f'^2\right)^{3/2}}\ .
\end{eqnarray}
Also the Ricci tensor and Ricci scalar are given by
\begin{eqnarray}\label{Ricci}
R_{\mu\nu}=\frac{2}{r^2}\left(
\begin{array}{ccc}
 -1 & 0 & 0 \\
 0 & \beta \left( \ln  \frac{r}{{L}}-\frac{1}{2}\right) & 1 \\
 0 & 1 & 0
\end{array}
\right),\hspace*{1cm}R=-\frac{6}{{L}^2}.
\end{eqnarray}
By making use of these expressions the entropy functional \eqref{EN} reads
\begin{eqnarray}\label{EE1}
S_A=\frac{2\pi{L}}{l_p} \int{\frac{dr}{r}\sqrt{1+f'^2+\frac{\beta}{2}f'^2\ln \frac{{L}}{r}}\left(1-2L^2 F\right)},
\end{eqnarray}
where
\begin{eqnarray}\label{Frb}
&&F=1+\frac{\beta}{1-\frac{\beta}{2} \ln \frac{r}{{L}}}-\frac{\beta}{1+\frac{\beta}{2} \ln \frac{r}{{L}}+f'^2}\cr &&\cr &&+\frac{1}{4\left(1-\frac{\beta}{2} \ln \frac{r}{{L}}\right)^2 \left(1+\frac{\beta}{2} \ln \frac{r}{{L}}+f'^2\right)^2}\times 
\bigg[\frac{ \beta^2\left(1+\ln \frac{r}{{L}} \left(1-\frac{\beta}{4}-\frac{\beta}{2} \ln \frac{r}{{L}}\right)\right)^2 f'^2}{1-\frac{\beta}{2} \ln \frac{r}{{L}}}\cr
&&\cr &&-\frac{4 \left(\left(1+\frac{\beta}{2}-\frac{\beta^2}{4} \ln ^2\frac{r}{{L}}\right) f'+\left(1+
\frac{\beta}{4}-\frac{\beta}{2} \ln \frac{r}{{L}}\right) f'^3-r \left(1-\frac{\beta^2}{4} \ln ^2\frac{r}{{L}}\right) f''\right)^2}{1+\frac{\beta}{2} \ln \frac{r}{{L}}+f'^2}\bigg].
\end{eqnarray}
Now the aim is to minimize $S_A$ to find a differential equation for $f$ whose solution is  the profile of the co-dimension two hypersurface in the bulk. To do so, we may consider the above 
expression as an action for the dynamical field $f$.  Since the action does not depend 
on $f$, the corresponding momentum is a constant of motion. It is then straightforward, to write down the equation of the conservation law, though in general it is not obvious whether the corresponding equation  can be solved exactly. Nevertheless since we are interested in small $\beta$ limit,
one may solve the equation perturbatively to find $f$. Indeed at leading order one finds
\begin{eqnarray}\label{bexpand}
f'(r)=\frac{r}{\sqrt{r_t^2-r^2}}\left(1+\frac{\beta}{4}\frac{r_t^2\ln\frac{r_t}{L}-r^2\ln\frac{r}{L}}{r_t^2-r^2}\right)+{\cal O}(\beta^2)\ ,
\end{eqnarray}
where $r_t$ is the turning point, which is related to the width of the entangling region $\ell$ via
the following constraint
\begin{eqnarray}\label{cons}
\ell=2\int_0^{r_t}drf'(r).
\end{eqnarray}
In particular at leading order one finds
\begin{eqnarray}
{\ell}=2r_t\left[1-\frac{\beta }{4}\left(1-\ln\frac{4r_t}{L}\right)+{\cal O}(\beta^2)\right]\ .
\end{eqnarray}
Putting everything together and keeping in our mind that at the leading order $\ell=2r_t$, one can evaluate   the holographic entanglement entropy as follows
\bea
S_{EE}&=&-\frac{\beta L}{4G_N}+\frac{\beta^2 L}{8G_N}\left[
\ln^2\left(\frac{\epsilon}{L}\right)-\frac{5}{6}\ln^2\left(\frac{\ell}{L}\right)+\frac{8}{9}\ln\left(\frac{\ell}{L}\right)+c_0\right]+{\cal O}(\beta^3)\cr &&\cr
&=&\frac{\beta b}{48}-\frac{\beta^2 b}{96}\left[
\ln^2\left(\frac{L}{\epsilon}\right)-\frac{5}{6}\ln^2\left(\frac{\ell}{L}\right)+\frac{8}{9}\ln\left(\frac{\ell}{L}\right)+c_0\right]+{\cal O}(\beta^3)\ ,
\eea
where $b=\frac{b_L+b_R}{2}$ is the new anomaly of the LCFT and $c_0$ is a numerical constant.

It is interesting to note that since the central charges of the NMG model at critical 
point are zero the leading universal log term associated to two dimensional CFT's, $\ln\frac{\ell}{\epsilon}$, is absent. Therefore  the
universal divergent term should come from higher order corrections to the entanglement 
entropy which has a new form. From dual field theory
point of view this corresponds to the fact that at the critical point the theory has new 
degrees of freedom. So, when we are computing the entanglement entropy, we are 
indeed measuring the entanglement between these new degrees of freedom. Therefore the 
short range behaviour of the entanglement entropy may be changed.
Moreover the coefficient of the universal term is given by the new anomaly.
This might indicate that the degrees of freedom of the new modes are controlled by the new anomaly. 



\section{Higher Dimensional Log-gravity}

In this section we will study  holographic entanglement entropy for certain
higher derivative $d+1$ dimensional gravity. The corresponding  action is 
\cite{Deser:2011xc}\footnote{This action 
can be found by setting $\gamma=0$ in the action given in\cite{Deser:2011xc} .}
\begin{equation}
I=-\frac{1}{2l_p^{d-1}}\int d^{d+1}x\,\sqrt{-g}\bigg[R-2\Lambda-\frac{1}{m^2}\left( R^{\mu\nu}R_{\mu\nu}
-\frac{d+1}{4d} R^{2}\right)\bigg].\label{eq:Quadratic_action}
\end{equation}
For generic values of the parameters $\Lambda$ and
$m$ the model has an AdS vacuum solution whose radius of the curvature, setting $x=-\frac{d(d-1)}{2L^2}$, can be obtained from  the roots of the following equation\cite{Deser:2011xc}
\be\label{Lam}
\frac{d-3}{2dm^2}\;x^2+x-\Lambda=0\ .
\ee
Here $L$ is the radius of the AdS solution. 

It can be shown that the model has a critical point where $m^2=\frac{(d-1)^2}{2L^2}$\cite{Deser:2011xc}.
 At this point the  corresponding equations of motion 
degenerate leading to a logarithmic solution. The aim of this section is to compute the 
entanglement entropy for this solution.

Holographic entanglement entropy for a generic higher derivative terms in a five dimensional  AdS solution
 has been studied in \cite{Bhattacharyya:2013gra}. Here we use the same procedure, though 
our main concern is the logarithmic solution.  In the present case following the  results of \cite{Fursaev:2013fta}, one needs to minimize the following entropy functional 
\be\label{EEd}
S_{EE}=\frac{2\pi}{l_p^{d-1}}\int d^{d-1}x\;\sqrt{g_{\rm ind}}\left\{1-\frac{1}{m^2}\left[\left(R_{\mu\nu}n^\mu_in^\nu_i-
\frac{1}{2}K_i^2\right)-\frac{d+1}{2d}R\right]\right\},
\ee
In what follows we will compute the entanglement entropy for the following  generic solution 
\be
ds^2=\frac{L^2}{r^2}\left[-\beta \ln\left(\frac{r}{L}\right) \;dx_+^2-2dx_+dx_-+d\rho^2+\rho^2 d\Omega_{d-3}^2+dr^2\right].
\ee
Note that in our computations we will consider  both $m$ and $\beta$ as free parameters, though 
one should keep in our mind that $\beta$ is zero when $m^2\neq (d-1)^2/2L^2$. In other 
words, for the logarithmic solution $m$ is fixed and is not a free parameter.
  
Let us consider  an entangling region in the shape of cylinder at fixed time and $0\leq \rho\leq \ell$. Then one needs to 
consider a co-dimension two
hypersurface in the bulk whose boundary coincides  with the boundary of the
entangling region. The corresponding hypersurface is given by
\be
 x_++x_-=0,\;\;\;\;\;\;\;\;\;\;\;\;\;\;\;\rho=f(r)\ .
\ee
 In this case the induced metric reads
\begin{eqnarray}\label{inducedd-dim}
ds^2=\frac{L^2}{r^2}\left((1+f'^2)dr^2+(2-\beta\log \frac{r}{L})dx_+^2+f^2d\Omega_{d-3}^2 \right)\ .
\end{eqnarray}
On the other hand two unit vectors normal to the co-dimension two  hypersurfaces are
\begin{eqnarray}\label{sigma1}
&&x_++x_-=\text{const.}\hspace*{1.36cm}n_1=\frac{L}{r \sqrt{2-\beta \log \frac{r}{L}}}(0,1,1,0,0,\cdots)\ ,\nonumber\\
&&\rho -f(r)=\text{const.}\hspace*{1.3cm}n_2=\frac{L}{r \sqrt{1+f'^2}}(-f',0,0,1,0,\cdots)\ .
\end{eqnarray}
It is then straightforward though tedious to compute the entropy functional. Indeed setting 
$F=2-\beta \ln \frac{r}{L}$ the  entropy functional \eqref{EEd} reads
\begin{eqnarray}\label{EEddim}
S_A&=&\frac{ 2\pi\Omega_{d-3}H_+L^{d-1}}{\sqrt{2}l_p^{d-1}}\int {dr}\;\frac{\sqrt{\left(1+f'^2\right)F}}{ r^{d-1}f^{3-d}} \;
\Bigg\{1-\frac{1}{m^2}\Bigg(\frac{(d-1)^2F+d\beta}{2L^2F}\\
&&-\frac{\bigg[\bigg(2(d-3) r F+ff'\bigg(2(d-1)F+\beta\bigg) \bigg) \left(1+f'^2\right)-2 r f f'' F\bigg]^2}{8L^2 f^2 F^2 \left(1+f'^2\right)^3}\Bigg)\Bigg\},\nonumber
\end{eqnarray}
where $\Omega_{d-3}$ is the volume of the  $S^{d-3}$ sphere  in the metric.

Now following our procedure in the previous section, one should  minimize the above entropy functional to find a differential equation for $f$. Again,
in general it is hard to solve the corresponding differential equation. Nevertheless one may solve the 
equation perturbatively in power of $\beta$ near the boundary. Indeed at leading order one finds
\be\label{tt}
f(r)=\ell-\frac{d-3}{2(d-2)}\frac{r^2}{\ell}+{\cal O}(\beta\; \&\; r^4).
\ee
We must emphasis that since we are solving the differential equation near the boundary, we can only 
extract information about the UV contributions or the most divergent terms of the entanglement 
entropy. Note also that the above equation is not valid for $d=3$. Indeed in this case the entangling region is a strip with the width $\ell$ and the equation of motion for $f$ at 
leading order can be exactly solved. More precisely one finds (see also\cite{RT:2006})
\be\label{tt3}
 f'(r)=\frac{r^2}{\sqrt{r_t^4-r^4}}+{\cal O}(\beta),
\ee
 where $r_t$ is the turning point which can be fixed in terms of $\ell$ by the constraint $\ell=
2\int_0^{r_t}dr f'(r)$.

Plugging the solution \eqref{tt} (or \eqref{tt3} for $d=3$) into the entropy functional \eqref{EEddim}, one can read the UV behaviour of the entanglement 
entropy. Indeed for $\beta=0$ where the solution is an AdS solution the 
most divergent term of the entanglement entropy for arbitrary $m$ is 
\be\label{D1}
S_{EE}=-\frac{2\pi \Omega_{d-3} L^{d-1}}{l_p^{d-1}}\;\left(1-\frac{(d-1)^2}{2m^2L^2}\right)
\;\Bigg[\frac{H_+\ell^{d-3}}{(d-2)\epsilon^{d-2}}+\cdots+{\tilde{c}_d}\frac{H_+}{\ell}\ln\frac{\epsilon}{\ell}\Bigg],
\ee
where $H_+$ is the height of the cylinder. Note that  the logarithmic term whose coefficient is a universal constant  is non-zero only for even $d$. More precisely,  one has
\be
{\tilde c}_3={\tilde c}_5=0,\;\;\;\;\;\;\;\;\;{\tilde c}_4=\frac{1}{8},\;\;\;\;\;\;\;\;{\tilde c}_6=-\frac{135}{2048}.
\ee
Since the entangling region is a cylinder, the coefficient of the universal part of the entanglement 
entropy is related to the central charge of the dual CFT\cite{Solodukhin:2008dh}. In the present case writing the universal part as $\frac{c_d}{3}\frac{H_+}{\ell}\ln\frac{\epsilon}{\ell}$, 
one finds\footnote{The case of $d=4$ has also been considered in \cite{Bhattacharyya:2013gra}.}
\be
c_4=\frac{3\pi^2 L^3}{2l_p^3} \left(1-\frac{9}{2m^2L^2}\right),\;\;\;\;\;\;\;\;\;\;\;
c_6=\frac{405\pi^3 L^5}{256l_p^5} \left(1-\frac{25}{2m^2L^2}\right).
\ee
From the above expressions it is evident that at the critical point where $m^2=\frac{(d-1)^2}{2L^2}$ 
all  divergent terms, including the logarithmic term (for even $d$), vanish. Of course  at this point the model exhibits a new logarithmic solution and therefore it is natural to study the entanglement entropy of this solution. In fact the situation is very similar to that in NMG case.  For this case, setting $m^2=\frac{(d-1)^2}{2L^2}$,  at leading order one finds
\be\label{D2}
S_{EE}=\beta \frac{d \pi \Omega_{d-3} L^{d-1}}{(d-1)^2l_p^{d-1}}
\;\Bigg[\frac{H_+\ell^{d-3}}{(d-2)\epsilon^{d-2}}+\cdots+{\tilde{b}_d}\frac{H_+}{\ell}\ln\frac{\epsilon}{\ell}\Bigg],
\ee
Again the logarithmic term is non-zero for even $d$. Indeed one has
\be
{\tilde b}_3={\tilde b}_5=0,\;\;\;\;\;\;\;\;\;{\tilde b}_4=\frac{1}{8},\;\;\;\;\;\;\;\;{\tilde b}_6=-\frac{315}{2048}.
\ee
Since at the critical point the dual theory is supposed to be a LCFT,  it is then natural to 
identify the universal constant with the new anomaly of the dual LCFT. More precisely writing
the universal term as $\frac{b_d}{48}\frac{H_+}{\ell}\ln\frac{\epsilon}{\ell}$,  we have
\be
b_4=-\frac{16\pi^2 L^3}{3l_p^3},\;\;\;\;\;\;\;\;\;\;\;
b_6=-\frac{567\pi^3 L^5}{80l_p^5} .
\ee

It is worth recalling that for higher dimensional CFT's there are several central charges and  depending 
on the shape of the entangling region the universal part of the entanglement entropy could be
proportional to different central charges. In particular in four dimensional CFT when the
entangling region is a sphere the universal part is proportional to $a$ while for cylinder
it is proportional to $c$ \cite{Solodukhin:2008dh}. On the other hand taking a strip as the 
entangling region there would be no universal part.

In this section, due to the symmetry of the logarithmic metric,  we have considered a cylinder entangling surface. So, we would expect to get a universal term in the expression of the 
entanglement entropy, at least for even dimensional CFT's. It seems natural to take the coefficient
of the universal part as a parameter of the dual LCFT.  Therefore  one may identify this 
parameter as the  new anomaly in the dual LCFT.

We  note, however,  that in the four dimensions, as we have already mentioned, the entangling 
region is essentially a strip. Thus,  the resultant entanglement entropy does not contain a 
universal part. Therefore from our computations one cannot read the new anomaly of the corresponding 
three dimensional LCFT, even if it is non-zero\cite{Johansson:2012fs}\footnote{We would like 
to thank T. Zojer for a comment on this point.}. Indeed, in this case,  what we have 
computed  is the coefficient of the most divergent term when the central charge is zero. Of course eventually
it might be related to the new anomaly.

As a final remark we note that in higher dimensions we could have also considered a co-dimension two hypersurface  which is parametrized as that in the equation \eqref{two}. In this case the entangling 
region is, indeed, a strip along the light like direction and the UV contribution to the 
entanglement entropy does not have a universal part, though the most divergent terms are still given 
by that in the equations \eqref{D1} and \eqref{D2}. One also observes that in this case
the divergent terms come from  the order of $\beta^2$, which is consistent with the NMG case.


\section{Discussions}

In this paper we have studied entanglement entropy of certain LCFT's by making use of their holographic 
descriptions. The gravity dual of the corresponding LCFT's may be provided by higher derivative gravities
 at  critical points where the corresponding equations of motion of the models degenerate leading to 
logarithmic  solutions which are not asymptotically locally AdS.

Our considerations are based on the following assumptions. Since so far there is no 
a systematic procedure to compute holographic entanglement entropy when the action has 
arbitrary higher derivative
terms, we have used the regularization of squashed cones method introduced in \cite{Fursaev:2013fta}.
In this method  one minimizes (extremizes) the corresponding entropy functional (see for example the equation
\eqref{EN}) to find the profile  of the co-dimension two
hypersurface in the bulk gravity. Then the entanglement entropy is given by the entropy functional
evaluated on this hypersurface.  We note, however, that  to compute the entanglement entropy, a priori, 
it is not obvious whether the entropy functional should be evaluated on the hypersurface which minimizes the whole entropy functional\footnote{We would like to thank the referee 
for his/her comment on this point.}.  Actually for using this procedure we are encouraged by the results of \cite{Bhattacharyya:2013gra} where the authors have shown that
the regularization procedure leads to the expected universal terms in the entanglement entropy for spherical and cylindrical entangling surfaces. 

We note that since there is no a proof for the above procedure, it is important to examine
how robust the results are. Indeed to see this, one may compute the entropy functional on a hypersurface which only minimizes the area. In this case one observes that although the finite terms of resultant entanglement entropy will be changed, the universal divergent terms remain unchanged. In other words it seems that 
the coefficient of the divergent terms are robust.

On the other hand the logarithmic solutions have a non-renormalizable mode which destroys 
the asymptotic behaviour from that of  an AdS solution.  Therefore it is not clear how to implement the holographic 
renormalization which is the main stone of AdS/CFT correspondence. Nevertheless it was
argued in \cite{Skenderis:2009nt} that this non-renormalizable mode should be associated to an
irrelevant operator in the boundary theory. Therefore for sufficiently small deformation one may still
use the CFT tools. Of course the results should only be trusted at leading order in the perturbation
in the coefficient of the irrelevant operator. 

Following this procedure we have considered an {\it ad hoc} parameter in front of the log term 
in the metric. Then we 
have computed the holographic entanglement entropy in leading order of the parameter. 
Therefore it is 
important to keep in our mind that what we have really computed is the entanglement entropy 
of a CFT which is slightly deformed by an irrelevant operator. So, the deformation parameter, $\beta$,
 appeared in the final expressions  of the entanglement entropy. Thus
one could only trust the results at the lowest  orders. Indeed if we had been able to directly 
apply AdS/CFT for log gravity, the parameter $\beta$ would have  not been appeared in the 
final expressions.

For a generic  two dimensional CFT the entanglement entropy has the following form
\be
S_{EE}=\frac{c_L+c_R}{6}\ln\frac{\ell}{\epsilon},
\ee 
On the other hand for the NMG model  one has $c_L+c_R=\frac{3L}{G_N}(1-\frac{1}{2m^2L^2})$
which shows that the $R^2$ terms contribute to the above universal term in the 
entanglement entropy. In particular at the critical point where $2m^2L^2=1$ it vanishes indicating 
the universal divergent term should  come from higher order terms in $\beta$. Indeed this 
is what we have found in this paper.

From dual field theory
point of view this corresponds to the fact that at the critical point the theory has new 
degrees of freedom. So, when we are computing the entanglement entropy, we are 
indeed measuring the entanglement between these new degrees of freedom. Therefore the 
short range behaviour of the entanglement entropy may be changed.
Moreover the coefficient of the universal term is given by the new anomaly.
This might indicate that the degrees of freedom of the new modes are controlled by the new 
anomaly (see also \cite{Cardy:2013rqg}).

In this paper we have also studied logarithmic CFT in higher dimensions  using 
 the higher dimensional logarithmic gravity. Actually the result  has a similarity with that  of  NMG
model. This, indeed, can be understood from  fact that in both cases the corresponding actions have the same form.

In the context of the  entanglement entropy the entropic c-function  may be defined by derivative of the entanglement entropy with respect to
the size of the entangling region\cite{Casini:2006es}
\be
c_{e}(\ell)=\ell\frac{\partial S_{EE}}{\partial \ell},
\ee
which is universal, positive, and due to strong sub-additivity property of entanglement 
entropy satisfies 
\be
\frac{\partial c_{e}(\ell)}{\partial\ell}\leq 0.
\ee
Although the NMG model at the critical point is not a  unitary theory and the c-function may not 
be applied, having found the corresponding entanglement entropy it might be useful to compute 
the entropic c-function for this model. Doing so, one finds that it is universal in the
sense that it does not depend on the UV cut off, though it is positive for particular values of the 
entangling region. Nevertheless its first derivative is negative which in turn indicates that the resultant 
entanglement entropy satisfies strong sub-additivity\cite{Casini:2006es}.

By making use of the holographic description of entanglement entropy, it would be interesting
to explore a possible analogue of c-theorem or perhaps b-theorem in LCFT's.  See  \cite{Cardy:2013rqg}
for recent discussions on this subject.

As a final remark let us make a comment on the holographic entanglement entropy of a two dimensional
LCFT whose gravity dual is provided by TMG at the critical point. The action of the TMG model may be
written as follows
\be
S_{TMG}=\frac{1}{16\pi G_N}\int d^3x \left[R+\frac{2}{L^2}+\frac{\epsilon^{\mu\nu\rho}}{4\mu}\left(R_{ab\;\mu\nu}\;\omega^{ab}_{\ ,\rho}+\frac{2}{3}\omega^a_{\ b,\mu}\omega^b_{\ c,\nu}\omega^c_{\ a,\rho}\right)\right]\ ,
\ee
where $\omega^a_{\ b,\mu}$ is the spin connection whose inner Lorentz indices
are denoted by $a,b,\cdots$ while the space-time indices are denoted by  $\mu ,\nu ,\cdots$. 

This model admits a logarithmic solution as \eqref{logm} at $\mu L=1$ while for a
 generic value of  $\mu$ the model has an AdS vacuum solution. It is conjectured that
the TMG model on an asymptotically locally AdS solution with a proper boundary condition 
would  provide a gravitational dual
for a two dimensional CFT with the following central charges
\be
c_L=\frac{3L}{2G_N}\left(1-\frac{1}{\mu L}\right),\;\;\;\;\;\;\;\;\;\;
c_R=\frac{3L}{2G_N}\left(1+\frac{1}{\mu L}\right).
\ee
Holographic entanglement entropy for this model has been studied in\cite{Sun:2008uf}
where it was shown that for AdS geometry the contribution of the Chern-Simons term vanishes. 

To compute the  holographic entanglement entropy, inspired by the results of \cite{Solodukhin:2005ah}
where the author has evaluated the contribution of the Chern-Simons term  to the entropy of BTZ black hole,
we  might naively  consider the following entropy functional
\begin{equation}\label{EECS}
S_A=\frac{1}{4 G_N}\int dx\;\left[\sqrt{g_{\rm ind}}+
\frac{L}{2}\;
 \omega_{ab,\mu}\ e^a_{\ \alpha}e^b_{\ \beta}\ \epsilon^{\mu\nu\alpha} (n_i^\alpha n^i_\mu)(n_j^\beta n^j_\nu)\right]\ ,
\end{equation}
where $e^a_{\ \alpha}$ is the vielbein and  the normal vectors $n_\mu^i$ are the same as those in section two\footnote{The non-zero components of the vielbeins and spin connections for the metric 
\eqref{logm} are
\bea
e^1_{\ r}=\frac{L}{r},\;\;\;\;e^2_{\ -}=\frac{L}{r},\;\;\;\;e^3_{\ +}=\frac{L}{r},
\;\;\;\;e^2_{\ +}=\frac{\beta L}{2r}\log(\frac{r}{L}),\;\;\;\;\;\;\;\;\;
\omega_{+,21}=
\frac{1}{r}\ ,\;\;\;\;\omega_{-,31}=
\frac{1}{r},
\;\;\;\;\;
\omega_{+,13}=
\frac{\beta}{2r}[1-\log(\frac{r}{L})]\ .\nn
\eea}. 
It is then straightforward to minimize the above entropy functional to find the  profile of the corresponding co-dimension two hypersurface and then the entanglement entropy. Doing so, for small $\beta$ at leading 
order one finds
\begin{equation}
S_{EE}=\frac{c_R+c_L}{6}\ln \frac{\ell}{\epsilon}+\frac{\beta b}{24}\ln 2+\CO\left(\beta^2\right)\ .
\end{equation}
Note that unlike the NMG model in the present case since $c_L+c_R$ is $\mu$ independent, the
universal part  remains unchanged, though one gets a correction due to logarithmic term in the action which 
is consistent with the results of \cite{Sun:2008uf}, namely setting $\beta=0$ the correction vanishes.

It is, however, important to note that although the result seems physically reasonable and also consistent
with the literature, it might be misleading\footnote{We would like to thank the referee for 
his/her comment on this point and in particular pointing out the importance of the extrinsic curvature.}. The reason is as follows.

Actually the equation \eqref{EECS} has been obtained in the context of the BTZ black hole where 
the extrinsic curvature vanishes, though in the present case where the general co-dimension two hypersurface might 
have non-zero  extrinsic curvature it is not clear how to modify the equation \eqref{EECS} to 
include the effects of the  extrinsic curvature. From the equation \eqref{tracek}  one observes that 
$K_1$ is non-zero and indeed is proportional to $\beta$.  Since we are interested in the  small $\beta$ 
limit, there might be a possibility to have  ${\cal O}(\beta)$ corrections to the entanglement entropy 
due to non-zero extrinsic curvature. It would be interesting to explore this possibility.

  
\section*{Acknowledgements}

We would like to thank  A. E. Mosaffa  for collaboration in the early stage of the paper and  useful discussions.
We also acknowledge the use of M. Headrick's excellent Mathematica package "diffgeo". We would like
to thank him for his generosity. We  would also like to thank the referees for their useful comments.



\begin{thebibliography}{99}
\baselineskip=8pt


\bibitem{RT:2006PRL}
S.~Ryu and T.~Takayanagi,
"Holographic Derivation of Entanglement Entropy from AdS/CFT,''
Phys. Rev. Lett. {\bf 96} (2006) 181602
[hep-th/0603001].

\bibitem{RT:2006}
S.~Ryu and T.~Takayanagi,
"Aspects of Holographic Entanglement Entropy,''
JHEP {\bf 0608} (2006) 045
[hep-th/0605073].

\bibitem{Gurarie:1993xq} 
  V.~Gurarie,
  ``Logarithmic operators in conformal field theory,''
  Nucl.\ Phys.\ B {\bf 410}, 535 (1993)
  [hep-th/9303160].

\bibitem{Deser:1981wh} 
  S.~Deser, R.~Jackiw and S.~Templeton,
  ``Topologically Massive Gauge Theories,''
  Annals Phys.\  {\bf 140}, 372 (1982)
  [Erratum-ibid.\  {\bf 185}, 406 (1988)]
  [Annals Phys.\  {\bf 185}, 406 (1988)]
  [Annals Phys.\  {\bf 281}, 409 (2000)].

\bibitem{Deser:1982vy} 
  S.~Deser, R.~Jackiw and S.~Templeton,
  ``Three-Dimensional Massive Gauge Theories,''
  Phys.\ Rev.\ Lett.\  {\bf 48}, 975 (1982).

\bibitem{Bergshoeff:2009hq} 
  E.~A.~Bergshoeff, O.~Hohm and P.~K.~Townsend,
  ``Massive Gravity in Three Dimensions,''
  Phys.\ Rev.\ Lett.\  {\bf 102}, 201301 (2009)
  [arXiv:0901.1766 [hep-th]].

\bibitem{Deser:2011xc}
  S.~Deser, H.~Liu, H.~Lu, C.~N.~Pope, T.~C.~Sisman and B.~Tekin,
  ``Critical Points of D-Dimensional Extended Gravities,''
  arXiv:1101.4009 [hep-th].

\bibitem{Alishahiha:2011}
M.~Alishahiha and R.~Fareghbal,
  ``D-Dimensional Log Gravity,''  Phys.\ Rev.\ D  {\bf 83}, 084052 (2011) 
[ arXiv:1101.5891 [hep-th]].

\bibitem{Gullu:2011sj}
  I.~Gullu, M.~Gurses, T.~C.~Sisman and B.~Tekin,
  ``AdS Waves as Exact Solutions to Quadratic Gravity,''
  arXiv:1102.1921 [hep-th].

\bibitem{Bergshoeff:2011ri}
  E.~A.~Bergshoeff, O.~Hohm, J.~Rosseel and P.~K.~Townsend,
  ``Modes of Log Gravity,''
  arXiv:1102.4091 [hep-th].

\bibitem{Grumiller:2008qz} 
  D.~Grumiller and N.~Johansson,
  ``Instability in cosmological topologically massive gravity at the chiral point,''
  JHEP {\bf 0807}, 134 (2008)
  [arXiv:0805.2610 [hep-th]].

\bibitem{AyonBeato:2004fq} 
  E.~Ayon-Beato and M.~Hassaine,
  ``pp waves of conformal gravity with self-interacting source,''
  Annals Phys.\  {\bf 317}, 175 (2005)
  [hep-th/0409150].

\bibitem{AyonBeato:2009yq} 
  E.~Ayon-Beato, G.~Giribet and M.~Hassaine,
  ``Bending AdS Waves with New Massive Gravity,''
  JHEP {\bf 0905}, 029 (2009)
  [arXiv:0904.0668 [hep-th]].

\bibitem{Grumiller:2008es} 
  D.~Grumiller and N.~Johansson,
  ``Consistent boundary conditions for cosmological topologically massive gravity at the chiral point,''
  Int.\ J.\ Mod.\ Phys.\ D {\bf 17}, 2367 (2009)
  [arXiv:0808.2575 [hep-th]].

\bibitem{Skenderis:2009nt} 
  K.~Skenderis, M.~Taylor and B.~C.~van Rees,
  ``Topologically Massive Gravity and the AdS/CFT Correspondence,''
  JHEP {\bf 0909}, 045 (2009)
  [arXiv:0906.4926 [hep-th]].

\bibitem{Alishahiha:2010bw} 
  M.~Alishahiha and A.~Naseh,
  ``Holographic renormalization of new massive gravity,''
  Phys.\ Rev.\ D {\bf 82}, 104043 (2010)
  [arXiv:1005.1544 [hep-th]].

\bibitem{Maldacena:1997re} 
  J.~M.~Maldacena,
 ``The Large N limit of superconformal field theories and supergravity,''
  Adv.\ Theor.\ Math.\ Phys.\  {\bf 2}, 231 (1998)
  [hep-th/9711200].

\bibitem{Grumiller:2009mw} 
  D.~Grumiller and I.~Sachs,
  ``AdS (3) / LCFT (2) - Correlators in Cosmological Topologically Massive Gravity,''
  JHEP {\bf 1003}, 012 (2010)
  [arXiv:0910.5241 [hep-th]].

\bibitem{Grumiller:2009sn} 
  D.~Grumiller and O.~Hohm,
  ``AdS(3)/LCFT(2): Correlators in New Massive Gravity,''
  Phys.\ Lett.\ B {\bf 686}, 264 (2010)
  [arXiv:0911.4274 [hep-th]].

\bibitem{Grumiller:2013at} 
  D.~Grumiller, W.~Riedler, J.~Rosseel and T.~Zojer,
  ``Holographic applications of logarithmic conformal field theories,''
  arXiv:1302.0280 [hep-th].

\bibitem{Wald:1993nt} 
  R.~M.~Wald,
  ``Black hole entropy is the Noether charge,''
  Phys.\ Rev.\ D {\bf 48}, 3427 (1993)
  [gr-qc/9307038].


\bibitem{Hung:2011xb} 
  L.~-Y.~Hung, R.~C.~Myers and M.~Smolkin,
  ``On Holographic Entanglement Entropy and Higher Curvature Gravity,''
  JHEP {\bf 1104}, 025 (2011)
  [arXiv:1101.5813 [hep-th]].

\bibitem{deBoer:2011wk} 
  J.~de Boer, M.~Kulaxizi and A.~Parnachev,
  ``Holographic Entanglement Entropy in Lovelock Gravities,''
  JHEP {\bf 1107}, 109 (2011)
  [arXiv:1101.5781 [hep-th]].

\bibitem{Fursaev:2013fta} 
  D.~V.~Fursaev, A.~Patrushev and S.~N.~Solodukhin,
  ``Distributional Geometry of Squashed Cones,''
  arXiv:1306.4000 [hep-th].

\bibitem{Bhattacharyya:2013gra} 
  A.~Bhattacharyya, M.~Sharma and A.~Sinha,
  ``On generalized gravitational entropy, squashed cones and holography,''
  arXiv:1308.5748 [hep-th].
  
\bibitem{Lewkowycz:2013nqa} 
  A.~Lewkowycz and J.~Maldacena,
  ``Generalized gravitational entropy,''
  JHEP {\bf 1308}, 090 (2013)
  [arXiv:1304.4926 [hep-th]].

\bibitem{Anninos:2013nja} 
  D.~Anninos, J.~Samani and E.~Shaghoulian,
  ``Warped Entanglement Entropy,''
  arXiv:1309.2579 [hep-th].

\bibitem{Bergshoeff:2009aq} 
  E.~A.~Bergshoeff, O.~Hohm and P.~K.~Townsend,
  ``More on Massive 3D Gravity,''
  Phys.\ Rev.\ D {\bf 79}, 124042 (2009)
  [arXiv:0905.1259 [hep-th]].

\bibitem{Liu:2009kc} 
  Y.~Liu and Y.~-W.~Sun,
  ``Consistent Boundary Conditions for New Massive Gravity in $AdS_3$,''
  JHEP {\bf 0905}, 039 (2009)
  [arXiv:0903.2933 [hep-th]].

\bibitem{Solodukhin:2008dh} 
  S.~N.~Solodukhin,
  ``Entanglement entropy, conformal invariance and extrinsic geometry,''
  Phys.\ Lett.\ B {\bf 665}, 305 (2008)
  [arXiv:0802.3117 [hep-th]].

\bibitem{Johansson:2012fs} 
  N.~Johansson, A.~Naseh and T.~Zojer,
  ``Holographic two-point functions for 4d log-gravity,''
  JHEP {\bf 1209}, 114 (2012)
  [arXiv:1205.5804 [hep-th]].

\bibitem{Cardy:2013rqg} 
  J.~Cardy,
  ``Logarithmic conformal field theories as limits of ordinary CFTs and some physical applications,''
  arXiv:1302.4279 [cond-mat.stat-mech].

\bibitem{Solodukhin:2005ah} 
  S.~N.~Solodukhin,
  ``Holography with gravitational Chern-Simons,''
  Phys.\ Rev.\ D {\bf 74}, 024015 (2006)
  [hep-th/0509148].
  
\bibitem{Sun:2008uf} 
  J.~-R.~Sun,
  ``Note on Chern-Simons Term Correction to Holographic Entanglement Entropy,''
  JHEP {\bf 0905}, 061 (2009)
  [arXiv:0810.0967 [hep-th]].


\bibitem{Casini:2006es} 
H.~Casini and M.~Huerta,
  ``A c-theorem for the entanglement entropy,''
  J.\ Phys.\ A {\bf 40}, 7031 (2007)
  [cond-mat/0610375].

\end{thebibliography}
\end{document}